\documentclass[aps,pre,preprint,superscriptaddress]{revtex4-1}  
\usepackage{amsmath,amsfonts,amssymb}
\usepackage{hyperref}
\usepackage{graphicx}
\usepackage{dcolumn}

\begin{document}
\title{Phase-Field Crystal Model with a Vapor Phase}

\author{Edwin J. Schwalbach}
\email{edwin.schwalbach@wpafb.af.mil}
\altaffiliation[Present Address: ]{Air Force Research Laboratory, Materials and Manufacturing Directorate, Wright Patterson Air Force Base, OH 45433, USA}

\author{James A. Warren}
\affiliation{Materials Science and Engineering Division, Material Measurement Laboratory, National Institute of Standards and Technology, Gaithersburg, MD 20899, USA }

\author{Kuo-An Wu}\email{kuoan@phys.nthu.edu.tw }
\affiliation{Department of Physics, National Tsing Hua University, Hsinchu 30013, Taiwan }

\author{Peter W. Voorhees}
\affiliation{Department of Materials Science and Engineering, Northwestern University, Evanston, IL 60208, USA}
\affiliation{Department of Engineering Sciences and Applied Mathematics, Northwestern University, Evanston, IL 60208, USA}


\begin{abstract}
Phase-Field Crystal (PFC) models are able to resolve atomic length scale features of materials during temporal evolution over diffusive time scales.  Traditional PFC models contain solid and liquid phases, however many important materials processing phenomena involve a vapor phase as well.  In this work, we add a vapor phase to an existing PFC model and show realistic interfacial phenomena near the triple point temperature.  For example, the PFC model exhibits density oscillations at liquid-vapor interfaces that compare favorably to data available for interfaces in metallic systems from both experiment and molecular dynamics simulations.  We also quantify the anisotropic solid-vapor surface energy for a 2D PFC hexagonal crystal and find well defined step energies from measurements on the faceted interfaces.  Additionally, the strain field beneath a stepped interface is characterized and shown to qualitatively reproduce predictions from continuum models, simulations, and experimental data.  Finally, we examine the dynamic case of step-flow growth of a crystal into a supersaturated vapor phase.  The ability to model such a wide range of surface and bulk defects makes this PFC model a useful tool to study processing techniques such as Chemical Vapor Deposition or Vapor-Liquid-Solid growth of nanowires.
\end{abstract}

\maketitle

\section{Introduction}
Phase-Field Crystal (PFC) models have emerged in recent years as a viable tool for simulating a broad array of phase transformations and materials processing phenomena\cite{Elder:2002,Elder:2004}. These models are similar to classic Phase-Field models in that they are based on a free energy functional of a continuously varying field\cite{Boettinger:2002}, the key difference being that the functional in a PFC model is constructed to have equilibrium states with a periodic lattice, in addition to the typical homogeneous states.  PFC models account for crystal properties and defects on atomic length scales and thus naturally include many crystallographic effects such as elasticity, anisotropic physical properties, and topological defects such as dislocations\cite{Elder:2002,Elder:2004,Berry:2006}.  Additionally, these models operate on diffusive time scales characteristic of phase transformations such as solidification.  Many technologically important processes have been studied by PFC including solid-solid phase transformations\cite{Greenwood:2010}, Kirkendall void formation\cite{Elder:2011}, eutectic solidification\cite{Greenwood:2011b}, and stress induced morphological instabilities of films\cite{Tegze:2009,Wu:2009,Elder:2010a} to name a few. 

Existing PFC models treat phase transformations between condensed phases such as liquid to crystal ($l \rightarrow c$) or solid state transformations between crystal structures ($c_\alpha \rightarrow c_\beta$)\cite{Elder:2004,Wu:2010a,Wu:2010b,Greenwood:2010,Greenwood:2011a}.  There are, however, many important processing pathways that involve a low density vapor phase, such as Chemical Vapor Deposition (CVD), as well as Vapor-Liquid-Solid (VLS)\cite{Wagner:1964} and Vapor-Solid-Solid (VSS)\cite{Kodambaka:2007,Wen:2009} nanowire growth.  Existing PFC models cannot currently simulate such phenomena because they do not include the vapor phase and therefore the critically important vapor-liquid and vapor-solid interfaces.  In some PFC studies, as in the case of Kirkendall void formation\cite{Elder:2011} and thin film morphological evolution\cite{Wu:2009}, a liquid phase has been used in place of a vapor.  However, the addition of a true vapor phase to the PFC model enables the study of problems with vapor-liquid-solid tri-junctions, a feature particularly important for VLS nanowire growth.  Such problems necessitate a model which captures the correct contact angles and wetting behavior of all three phases\cite{Roper:2010,Schwalbach:2011,Schwalbach:2012}.  

In this work, we describe an extension to previous PFC models that incorporates a low density vapor phase to enable simulations of the above phenomena.  A vapor phase has a significantly lower density than either a liquid or solid, and atoms in the vapor are essentially electronically isolated due to this relatively large atomic spacing.  The contribution to the vapor's free energy from atomic correlations is then negligible.  In order to model a transition between the highly correlated condensed phases and a low density vapor, we introduce an order parameter $\eta$ that changes smoothly from 1 to 0 between the vapor and condensed phases.  This order parameter modulates a direct correlation function $C_2$ of the type used by Greenwood et al.~\cite{Greenwood:2010,Greenwood:2011a}, which is an important contributor to the free energy of the condensed phases, but is negligible in the vapor phase.  We focus on pure materials in the present work, however we note that the addition of an alloying component\cite{Elder:2007,Greenwood:2011b} to the present model is necessary to study a process such as VLS nanowire growth.

Another approach to developing a PFC model with a vapor phase is to add another potential well to the traditional PFC free energy function near zero density that has a vanishingly small contribution from two-body correlations.  This would result in a system with a mean field critical point between the liquid and vapor phases similar to a van der Waals fluid\cite{Plischke:2006}, with a vapor phase that has no significant contributions from two-body correlations.  This approach is computationally attractive as it does not require the addition of a new field, $\eta$, and its accompanying evolution equation.  However, we believe the simplicity and control offered by the two-field approach and the flexibility of the associated independently controlled parameters outweigh this benefit, particularly at temperatures and densities away from the liquid-vapor critical point.

In the following work, we show that this model reproduces many important physical behaviors.  First, the equilibrium density-temperature phase diagram exhibits common features for a pure material such as a triple point, and we can examine coexistence and drive phase transformations between the various states by changing the system's temperature through the strength of $C_2$.  The new PFC model has the advantage that it can treat a range of surface phenomena realistically.  The interface between the homogeneous liquid and vapor phases exhibits significant structure with density ordering in the liquid similar to that observed in molecular dynamics (MD) simulations of lithium, magnesium, and aluminum~\cite{Gonzalez:2007}.  We quantify this effect and show that the model can be parametrized to quantitatively agree with experimental measurements of density oscillations in liquid-vapor Gallium interfaces\cite{Regan:1995,Regan:1997}.  Also, we demonstrate that 2D solid-vapor interfaces are strongly anisotropic and have well defined step energies that are a function of facet orientation.  Additionally, we show several examples of steps on the facets of a body centered cubic (BCC) crystal.  We find that the elastic strain field beneath a step is qualitatively consistent with results obtained for aluminum and nickel surfaces~\cite{Chen:1986,Chen:1989}.  Finally, we examine dynamic behavior by driving the system with a mass source and observe step-flow growth of solid-vapor interfaces, indicating that the PFC model could be used to simulate growth processes such as CVD.  The appendices and supplemental material include some of the finer points of our numerical and analytical techniques.

The remainder of this work is organized as follows.  In Sec.~\ref{sec:Model} we describe the PFC model in detail including the free-energy functional.  In Sec.~\ref{sec:Results}, we describe the bulk phase diagram, and properties of liquid-vapor and solid-vapor interfaces as well as describing step-flow growth of solids.   Where possible, we compare the PFC model results to other simulations or experiments.  Finally, in Sec.~\ref{sec:Conclusions} we make concluding remarks and suggestions for future work.

\section{Model}
\label{sec:Model}
In this section, we describe a PFC model of the solid, liquid, and vapor states.  We develop the free energy functional for the model and then give the evolution equations to be employed in Sec.~\ref{sec:Results}.  Additionally, we describe the process for computing the equilibrium phase diagram.

\subsection{Free Energy Functional}
We characterize the system with a spatially varying atomic density probability $\rho\left( \vec{r} \right)$ with position vector $\vec{r}$.  This model describes a pure material that exhibits three phases: a crystalline solid that has spatially varying atomic density with lattice symmetry $c$ and mean atomic density $\bar{\rho}_s$, a liquid phase with homogeneous density $\bar{\rho}_l$, and finally a vapor phase with homogeneous density $\bar{\rho}_v$.  The actual state exhibited by a given system depends on the temperature $T$ and mean density of the whole system $\bar{\rho}$, and all three phases can only be in simultaneous equilibrium at the triple point temperature $T_{tr}$.  We introduce an order parameter $\eta\left( \vec{r} \right)$ which takes on a value of 1 in the vapor phase, and 0 in the condensed phases.  A convenient non-dimensional scaled density $\psi$ is
\begin{equation}
	\psi = \left(\rho - \rho_0\right) / \rho_0,
\end{equation}
where $\rho_0$ is a reference density to be described shortly.

The total free energy of the system $\mathcal{F}$ in domain $\mathcal{V}$ is expressed as a functional of $\rho$ and $\eta$,
\begin{equation}
	\mathcal{F} = \rho_0 k_B T F = \rho_0 k_B T \int_\mathcal{V} f\left[ \rho\left( \vec{r} \right), \eta\left( \vec{r} \right), \nabla \eta\left( \vec{r} \right) \right] d\vec{r},
	\label{eqn:F_functional_basic}
\end{equation}
where the non-dimensional free energy density $f$ is
\begin{equation}
	f = g\left(\eta \right)f_v\left( \rho \right) + \left[1 - g\left(\eta \right) \right] f_{pfc}\left( \rho \right) + W \, h\left(\eta\right) + \frac{\kappa}{2} \left| \nabla \eta \right|^2,
	\label{eqn:F_functional}
\end{equation}
and $f_v\left( \rho \right)$ and $f_{pfc}\left( \rho \right)$ are the free energies densities of the vapor and condensed phases respectively, $g$ is  an interpolating function, and $h$ is a barrier function.  The explicit $\vec{r}$ dependence of the fields $\rho$ and $\eta$ has been dropped for clarity.  The parameters $W$ and $\kappa$ are an energy barrier and gradient energy coefficient respectively.  In this work, we employ the common polynomial interpolating function $g\left(\eta\right)$ and barrier function $h\left(\eta \right)$~\cite{Boettinger:2002}:
\begin{align}
	g \left( \eta \right) =& \eta^3 \left(6\eta^2 - 15\eta + 10 \right),\\
	h \left( \eta \right) =& \eta^2 \left( \eta - 1 \right)^2.
\end{align}

We use the model of Greenwood et al.~for the condensed phase free energy density $f_{pfc}$ to allow flexibility to control the crystal structure\cite{Greenwood:2010,Greenwood:2011a}.  This free energy is most simply expressed in terms of the scaled density $\psi$, 
\begin{equation}
	f_{pfc} = \frac{\psi^2}{2} - \nu \frac{ \psi^3}{6} + \xi \frac{\psi^4}{12} - \frac{\psi}{2} C_2 * \psi,
\end{equation}
where the convolution $C_2 * \psi$ is
\begin{equation}
	C_2 * \psi = \int_\mathcal{V} C_2\left( \left| \vec{r} - \vec{r}' \right| \right)  \psi\left(\vec{r}'\right) d\vec{r}'.
	\label{eqn:convolution_definition}
\end{equation}
The polynomial terms are an expansion of an ideal gas about $\rho_0$, and the coefficients $\nu$ and $\xi$ allow for deviations from the ideal behavior.  The quantity $C_2\left(\left| \vec{r} - \vec{r}' \right| \right)$ is the direct two-body correlation function, and from now on we will  refer to it as $C_2$ for brevity.  $C_2$ is engineered to have peaks in reciprocal space for wave vectors that are characteristic of the desired crystal lattice as described in Refs.~\citenum{Greenwood:2010} and \citenum{Greenwood:2011a}.  The Fourier transform of $C_2$ is denoted $\hat{C}_2\left( k \right)$ and is constructed using $N$ Gaussian peaks in $k$-space according to the procedure described in Ref~\citenum{Greenwood:2011a}.  Each peak has the form
\begin{equation}
	\hat{C}_{2,i} \left( k \right) = \exp \left( - \left(T/T_{0}\right)^2 \right) \exp \left( - \left( k - k_i \right)^2/ \left(2 {\alpha_i}^2\right)  \right)
	\label{eqn:CorrelationFunction}
\end{equation}
for $i=1,2,...,N$ and $k = \left|\vec{k}\right|$.  The peak locations $k_i$ and widths $\alpha_i$ for $i > 0$ control the crystal structure, anisotropy, and defect energies, and the temperature scale $T_{0}$ sets the peak amplitude.  In principle, each peak could have its own scale $T_{0,i}$, but in this work we will focus on systems with $N=1$.  In addition to peaks with positive amplitude at $k_i>0$, a peak with negative amplitude $A_0$ at $k_0=0$ can also be included.  A detailed discussion of this quantity as well as how to construct $C_2$ to achieve specific crystal structures is contained in Ref.~\citenum{Greenwood:2011a}.

If the amplitude of the first peak $\hat{C}_2\left( k_1 \right)$ is known at a given temperature $T_\text{ref}$, then the temperature scale $T_{0}$ is set according to the relation
\begin{equation}
	T_0 = T_\text{ref} \left[ - \ln \hat{C}_2 \left( k_1 \right) \right]^{-1/2}.
	\label{eqn:tempScale}
\end{equation}
Additionally, the peak width $\alpha_1$ is related to both the peak amplitude and the second derivative of the correlation function with respect to $k$ evaluated at the first peak $\hat{C}_2''\left(k_1\right)$,
\begin{equation}
	\alpha_1 = \sqrt{-\hat{C}_2 \left( k_1 \right) / \hat{C}_2''\left( k_1 \right)},
	\label{eqn:alphaScale}
\end{equation}
which we assume to be temperature independent in this work.  The combination of Eqs.~\ref{eqn:tempScale} and \ref{eqn:alphaScale} allow us to parameterize the PFC model using information for $\hat{C}_2$ determined either experimentally or via MD simulations.  Finally, we point out that $\hat{C}_2$ is related to the experimentally accessible structure factor $S\left( k \right)$ through~\cite{Chaikin:1995}
\begin{equation}
	\hat{C}_2\left( k \right) = 1 - S\left( k \right)^{-1}.
	\label{eqn:Ck_Sk_relation}
\end{equation}

The vapor phase free energy density $f_v$ is modeled as a simple quadratic well,
\begin{equation}
	f_v =\frac{b}{2}\left( \psi - \psi^0_v \right)^2 + \Delta.
	\label{eqn:f_v}
\end{equation}
The parameter $b$ controls the width of the well and therefore the bulk modulus of the vapor phase, and the parameter $\Delta$ sets the energy of the vapor phase with respect to the condensed reference state and is used to control aspects of the phase diagram including $T_{tr}$.  The energy well is centered at $\psi^0_v$, and this parameter can therefore be used to adjust the density of the equilibrium vapor.  In general, $b$, $\psi_v^0$, and $\Delta$ are all functions of $T$, however, for simplicity we take them to be constants.  Also, a more complex dependence on the density (e.g., logarithmic) could be employed for $f_v$, but this typically incurs a computational cost.

For convenience, we define the mean density $\bar{\rho}_\beta$ of a phase $\beta$ as
\begin{equation}
	\bar{\rho}_\beta = \frac{\rho_0}{\mathcal{V}_\beta} \int_{\mathcal{V}_\beta} \left(1 + \psi\left( \vec{r} \right) \right)d\vec{r},
	\label{eqn:meanDensity}
\end{equation}
where $\mathcal{V}_\beta$ is a volume containing only phase $\beta$.  For the homogeneous liquid and vapor phases, the equilibrium density is spatially uniform, and thus $\rho\left( \vec{r} \right)$ is $\bar{\rho}_l$ or $\bar{\rho}_v$ respectively.  However, in the crystalline phase, the integral over the equilibrium $\rho\left( \vec{r} \right)$ field is non-trivial, and must be performed in order to evaluate $\bar{\rho}_s$.  In practice, we employ a finite impulse response (FIR) filter to smooth the density before computing the mean value to ensure that the resulting mean density is independent of the extent of $\mathcal{V}_s$ as has been done for other PFC and MD simulations\cite{Davidchack:1998,Tegze:2009,Tegze:2011}.  Similarly, we define the non-dimensional mean free energy density $\bar{f}_\beta$ of phase $\beta$ as
\begin{equation}
	\bar{f}_\beta \left( \bar{\rho}_\beta \right) = \frac{F}{\mathcal{V}_{\beta}},
	\label{eqn:MeanEnergyDensityDfn}
\end{equation}
where $F$ is given by Eq.~\ref{eqn:F_functional_basic}.  This expression simplifies considerably for the vapor and liquid, but the integral remains when evaluating the solid, i.e.,
\begin{align}
	\bar{f}_v \left( \bar{\psi}_v \right) &= f_v\left( \bar{\psi}_v \right),		\\
	\bar{f}_l \left( \bar{\psi}_l \right) &= \left[ 1 - A_0 \right]\frac{{\bar{\psi}_l}^2}{2} - \nu \frac{ {\bar{\psi}_l}^3}{6} + \xi \frac{{\bar{\psi}_l}^4}{12},\\
	\bar{f}_s \left( \bar{\psi}_s \right) &= \frac{1}{\mathcal{V}_s}\int_{\mathcal{V}_s} f_{pfc}\left( \psi_s \left( \vec{r} \right) \right)d\vec{r},\label{eqn:meanSolidFreeEnergy}
\end{align}
where $\psi_s \left( \vec{r} \right)$ in Eq.~\ref{eqn:meanSolidFreeEnergy} is the equilibrium crystal density profile and $A_0$ is the amplitude of $\hat{C}_2$ at $k=0$.

\subsection{Evolution Equations}
We assume the system is isothermal, so both equilibrium and dynamics can be most simply considered using the Helmholtz Free energy.  We postulate that the conserved field $\rho$ evolves according to diffusive dynamics, and that the non-conserved $\eta$ field evolves via an Allen-Cahn equation.  For simplicity, we assume spatially uniform density and order parameter mobilities, $M_\psi$ and $M_\eta $, respectively.  In terms of the non-dimensional scaled density, evolution is described by
\begin{align}
	\frac{\partial \psi}{\partial t} =&  M_\psi \nabla^2 \frac{\delta F}{\delta \psi}, \label{eqn:psiEvo} \\
	\frac{\partial \eta}{\partial t} =& -M_\eta         \frac{\delta F}{\delta \eta}.  \label{eqn:etaEvo}
\end{align}
For completeness, from Eq.~\ref{eqn:F_functional}, we have
\begin{align}
	\frac{\delta F}{\delta \psi} &=g \, b \left( \psi - \psi_v^0 \right) + \left(1 - g \right)\left[ \psi - \frac{\psi^2}{2} + \frac{\psi^3}{3}\right]\notag \\
	& + \left( \frac{g}{2} - 1 \right) \left( C_2 * \psi \right) +\frac{1}{2} C_2 * \left( g \psi \right) .\\
	\frac{\delta F}{\delta \eta}&= 30 \, h \left[f_v - f_{pfc} \right] + 2 W \eta\left(2 \eta - 1\right)\left(\eta - 1 \right) - \kappa \nabla^2 \eta .
\end{align}
This system of equations is evolved in time according to a semi-implicit Fourier spectral scheme described in Appendix~\ref{sec:Numerics}.  The quantity $M_\psi$ is set equal to the self-diffusion coefficient $D$ in the solid which results in the correct time scales for diffusion of the mean density.  We also take the magnitude of $M_\eta$ sufficiently large with respect to $M_\psi$ to ensure that $\eta$ evolution is rapid compared to the relatively slow process of mass diffusion.

\subsection{Equilibrium}
\label{sec:equibTheory}
For an isothermal system, two homogeneous phases $\alpha$ and $\beta$ are in equilibrium when they have uniform chemical potential $\mu$ and pressure (in the case of a planar interface),
\begin{align}
	\frac{\partial \bar{f}_\alpha}{\partial \psi} &= \frac{\partial \bar{f}_\beta}{\partial \psi}  = \mu, \label{eqn:equal_mu}\\
	\bar{f}_\beta &= \bar{f}_\alpha + \mu \left(\bar{\psi}_\beta - \bar{\psi}_\alpha \right). \label{eqn:equal_p}
\end{align}
This system of equations is a common tangent construction which can be solved for the coexistence densities $\bar{\psi}_\alpha$ and $\bar{\psi}_\beta$ at a given temperature.

We have simple polynomial expressions for $\bar{f}_v$ and $\bar{f}_l$ which can be used in Eqs.~\ref{eqn:equal_mu} and \ref{eqn:equal_p} directly.  However, for the crystalline phase the free energy depends on the spatially varying density $\psi\left( \vec{r} \right)$ as described in Eq.~\ref{eqn:meanSolidFreeEnergy}.  There are one- or two-mode approximations for some simple crystal structures that can be used to estimate equilibrium $\psi_s\left( \vec{r} \right)$ and therefore $\bar{f}_s\left( \bar{\psi}_s \right)$\cite{Elder:2004,Greenwood:2011a}.  However, for complicated crystal structures, accurate approximations can require many terms, and numeric minimization of the energy with respect to the amplitude of each mode is necessary.  Instead, we determine $\bar{\psi}_s$ and $\bar{f}_s$ numerically by equilibrating a series of single phase periodic systems to find $\psi_s\left( \vec{r} \right)$ over a range of $T$ and $\bar{\psi}_s$, and then evaluate the integrals in Eqs.~\ref{eqn:meanSolidFreeEnergy} numerically.  For each value of $T$, a quadratic polynomial is fit to the numeric data for $\bar{f}_s \left( \bar{\psi}_s \right)$ for densities near the equilibrium value, and this approximation is then used in Eqs.~\ref{eqn:equal_mu} and \ref{eqn:equal_p}.  Also, for the parameters chosen in this work, we find that the free energy is minimized when the lattice parameter is equal to the value prediced by the a perfect lattice given the value of $q_1$.

\section{Results and Discussion}
\label{sec:Results}

In this section, we consider some implications of the model described in Sec.~\ref{sec:Model}.  We first compute the density - temperature phase diagram for a system with vapor, liquid, and BCC solid phase.  Then, we discuss the structure of both liquid-vapor and solid-vapor interfaces.  Finally, we consider the dynamic case of step-flow growth.

\subsection{Numerical Calculation of Phase Equilibrium}
\label{sec:equib}
In this section, we choose parameters to produce a system with a triple point at $T=1811$~K and solid BCC phase with a lattice parameter $a_0 = 0.298$~nm based on Fe and use a correlation function with only one peak.  The height, second derivative, and position of the peak in $\hat{C}_2$ are taken from Ref.~\citenum{Wu:2007}, and Eqs.~\ref{eqn:tempScale} and \ref{eqn:alphaScale} are then used to set $T_0$ and $\alpha_1$.  The solid-liquid density difference is controlled with $A_0$\cite{Greenwood:2011a}, and the amplitude of the density waves and the magnitude of the solid-liquid surface energy are further adjusted with the parameters $\nu$ and $\xi$.  The parameters $b, \psi_v^0$, and $\Delta$ are adjusted to set the triple point at the desired temperature.  Finally, the reference density $\rho_0$ is chosen to approximate the liquid density of Fe at the triple point.  A summary of all simulation parameters for this section is given in Table~\ref{tbl:BCCParams}.

\begin{table}
\caption{Model parameters used for the BCC + Liquid + Vapor system.  The values $k_1$, $\hat{C}_2\left( k_1 \right)$, and $\hat{C}_2''\left( k_1 \right)$ are taken from Ref.~\citenum{Wu:2007} are based on MD simulations of liquid Fe at 1820~K.  The mobility $M_\rho$ is based on the self diffusivity in $\delta$-Fe near its melting point reported in Ref.~\citenum{James:1966}}{}
\label{tbl:BCCParams}
\begin{ruledtabular}
\begin{tabular}{l d c}
	Quantity 				& \text{Value}		 & \text{Unit} 		\\ \hline
	$k_1$					& 2.985 \times 10^{10}&$\text{m}^{-1}$	\\
	$\rho_0$				& 6.57 \times 10^{28}&$\text{m}^{-3}$	\\
	$\hat{C}_2\left( k_1 \right)$	& 0.67		 & - 				\\
	$\hat{C}_2''\left( k_1 \right)$	& -10.4  \times 10^{-20}	&$\text{m}^2$  \\
	$\alpha_0$				& 2 \alpha_1		&$\text{m}^{-1}$	\\
	$\alpha_1$				& 2.53 \times 10^{9}&$\text{m}^{-1}$	\\
	$A_0$					& -1.				& -					\\
	$T_0$					& 2865.				&$\text{K}$ 		\\
	$\psi_v^0$				& -1.238			& -					\\	
	$\Delta$				& -0.2920			& -					\\	
	$\nu$					& 0.5376			& -					\\
	$\xi$					& 0.1				& -					\\
	$b$						& 1.0				& -					\\
	$W$						& 10.				& -					\\
	$\kappa$				& 6.25\times 10^{-21}&$\text{m}^2$		\\
	$M_\rho$				& 10^{-11}			&$\text{m}^2 \cdot \text{s}^{-1}$	\\
	$M_\eta$				& 10^{11}			&$\text{s}^{-1}$ 	\\	
	$\Delta t$				& 4.0 \times 10^{-10}&$\text{s}$		\\
	$\Delta x$				& 1.86 \times 10^{-11}&$\text{m}$
\end{tabular}
\end{ruledtabular}
\end{table}

\begin{figure}
\includegraphics{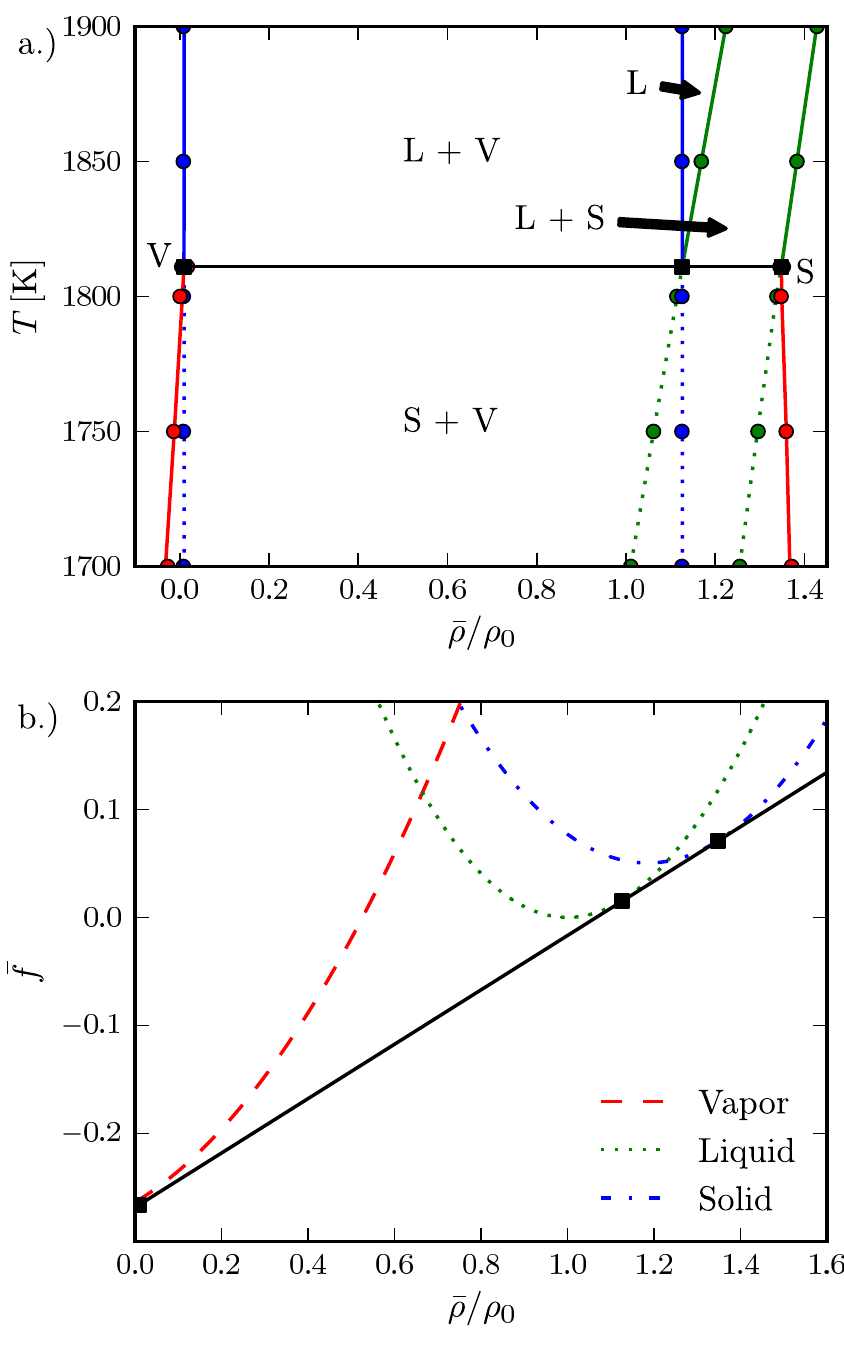}
\caption{(Color online) a.)  Stable (solid lines) and metastable (dotted lines) phase boundaries.  Measured coexistence densities for two-phase (circles) and three phase (solid squares) simulations agree well with the theoretical predictions.  b.) Free energy densities for each phase at the triple point $T_\text{tr}=1811$~K and the common tangent line.}
\label{fig:phase-diagram}
\end{figure}

Figure~\ref{fig:phase-diagram}a.) shows the phase diagram in $\bar{\rho} - T$ space using free energy parameters in Table~\ref{tbl:BCCParams}, where the densities are normalized by $\rho_0$.  The solid lines in the diagram are computed according to the procedure described in Sec.~\ref{sec:equibTheory}, and dotted lines, calculated in the same fashion, are metastable states at temperatures below $T_{tr}$.  Solid circles indicate coexistence densities measured from simulations which exhibit planar two-phase coexistence between liquid-vapor or liquid-solid phases.  These simulations are carried out with periodic boundary conditions and initial conditions with sharp interfaces, and are relaxed until the chemical potential is uniform.  The initial condition for the solid region is based on a one-mode sinusoidal approximation of a BCC crystal.  Simulation domains have a length normal to the interface $l_z \geq 96a_0$, and three-dimensional liquid-solid simulation domains have dimensions in the plane of the interface of $\sqrt{2} a_0 \times a_0$.  Densities are numerically measured in regions away from the interfaces after smoothing with an FIR filter.  The resulting measured two-phase coexistence densities agree well with the estimated phase boundaries as shown in Fig.~\ref{fig:phase-diagram}.

The system exhibits a triple point at $T_{tr} = 1811$~K with $\rho_v \approx \rho_l / 100$ and a solid-liquid density difference of 16.5~\% of $\bar{\rho}_s$.  Figure~\ref{fig:phase-diagram}b.) shows the free energy densities of each phase at $T_{tr}$ along with a common tangent line and equilibrium densities.  Note that the phase diagram exhibits the basic characteristics of a typical pure material near $T_{tr}$: low temperature solid-vapor coexistence, solid-liquid coexistence at higher temperatures and densities, and liquid-vapor coexistence at high temperatures and low mean densities.

For $T > T_{tr}$, the liquid-vapor equilibrium phase boundaries are vertical because the parameters $\psi_v^0$ and $\Delta$ in Eq.~\ref{eqn:f_v} are assumed to be temperature independent.  There is typically a critical point between the liquid and vapor phases at elevated temperature, and thus the phase boundaries should slope towards each other.  While the present PFC model does not produce a critical point, the phase boundary slopes could, in principal, be adjusted by using temperature dependent $\psi_v^0$ and $\Delta$, although we have not attempted this.  Also, because of its lack of critical point between the liquid and vapor phases, this model is most suitable for examining processing conditions near the triple point.

\begin{figure}
\includegraphics{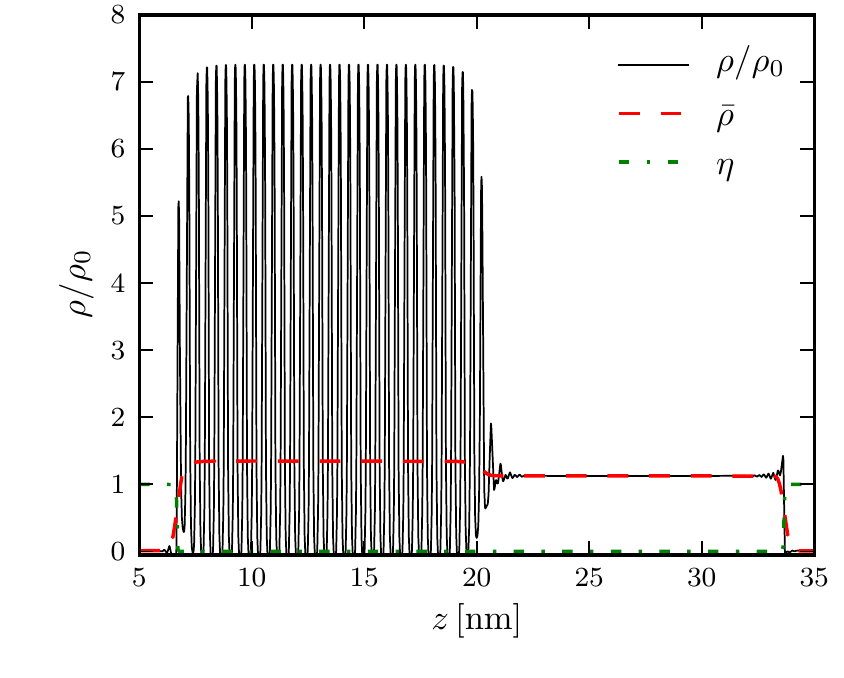}
\caption{(Color online)  Equilibrium density profile along the $z$ direction for the three-phase simulation at the triple point $T = 1811$~K shown in Fig.~\ref{fig:phase-diagram}.  A portion of the vapor phase is not shown for clarity.}
\label{fig:threePhaseSystem-trace}
\end{figure}

To test the numeric estimate of the invariant reaction at $T_{tr}$, a system with $\bar{ \rho } = 0.83$ and $T = 1811$~K was set up with an initial condition with slabs of BCC crystal, liquid, and vapor each occupying one third of the totald domain as shown in Fig.~\ref{fig:threePhaseSystem-trace}a.).  The domain has dimensions $l_x \times l_y \times l_z = a_0\sqrt{2} \times a_0 \times 192a_0$. The mean densities of each phase in the initial condition were chosen according to the invariant reaction in the phase diagram Fig.~\ref{fig:phase-diagram}(squares).  This system was numerically relaxed over the characteristic mass diffusion time $t_\rho = {l_z}^2 /  \left(4 M_\rho \right)$ where $l_z$ is the system size normal to the interfaces.  

The volume fraction of each phase was essentially unchanged during the relaxation time, indicating that the system is indeed at the triple point.  Additionally, the measured mean densities for each of the three phases remained within 1~\% of the values indicated in the phase diagram (Fig.~\ref{fig:phase-diagram}b.), squares).  Cooling this system to 5~K below $T_{tr}$ resulted in solid growth into the liquid phase, and heating the system to 5~K above $T_{tr}$ induced melting of the solid.  The numeric simulations are in good agreement with the temperature and density estimates for the invariant reaction described in Fig.~\ref{fig:phase-diagram}.  Additionally, Fig.~\ref{fig:threePhaseSystem-trace}b.) indicates that there is significant structure in the liquid near the liquid-vapor interface which will be explored in Sec.~\ref{sec:liquid-vapor}.

In addition to three-phase equilibrium, the relative surface energies of the liquid-solid, liquid-vapor and solid-vapor interfaces are of interest for cases where triple junctions play an important role in evolution.  In particular, the Young's angle $\theta_Y$, defined
\begin{equation}
	\cos \theta_Y = \left( \gamma_{sv} - \gamma_{sl} \right) / \gamma_{lv},
	\label{eqn:YoungsAngle}
\end{equation}
is important for scenarios such as VLS nanowire growth\cite{Roper:2010,Schwalbach:2011,Schwalbach:2012}.  We use equilibrium density and order parameter profiles for two phase eqiuilbrium interfaces to evaluate the surface energy numerically according to the procedure outlined in Refs.~\citenum{Wu:2007} and \citenum{Oettel:2012}.  Details are given in appendix~\ref{sec:SurfaceEnergy}.  Table~\ref{tbl:BCCgammas} shows the numerically computed surface energies for each of the three types of interface at the triple point.  Both the solid-liquid and solid-vapor interfacial energies are anisotropic, and we report values for $\left(110\right)$ and $\left(100\right)$ type interfaces as a representative range.
\begin{table}
\caption{Measured surface energies at 1811~K using simulation parameters in Table~\ref{tbl:BCCParams}.  Ranges are given for the anisotropic solid-liquid and solid-vapor energies.}
\label{tbl:BCCgammas}
\begin{ruledtabular}
\begin{tabular}{l d}
	Interface 		& \gamma \left[ \text{J} \cdot \text{m}^{-2} \right] \\ \hline
	Liquid-Vapor	& 0.0946\\
	Solid-Liquid	& 0.0516 \text{ to } 0.0527 \\
	Solid-Vapor		& 0.1388 \text{ to } 0.1478
\end{tabular}
\end{ruledtabular}
\end{table}

Based on the behavior of the classic phase-field model, we expect $\gamma_{sv}$ and $\gamma_{lv}$ to be approximately linearly proportional to $\sqrt{\kappa W}$, while the interface thickness is proportional to $\sqrt{\kappa / W}$~\cite{Boettinger:2002}.  However, $\gamma_{sl}$ is independent of both $\kappa$ and $W$ as the $\eta$ field is uniformly zero in this interface.  Equation~\ref{eqn:YoungsAngle} then suggests that $\theta_Y$ can be controlled by modifying the factor $\sqrt{\kappa W}$.  To test this, we simulate two phase liquid-vapor and solid-vapor systems with parameters given in Table~\ref{tbl:BCCParams}, except we multiply both $\kappa$ and $W $ by a factor $\chi$.  Note that both $\kappa$ and $W$ are modified by the same factor in order to keep the interface thickness approximately constant.  The values $\gamma_{sv}$ and $\gamma_{lv}$ are measured from two-phase numeric simulations over a range of $\chi$ values, and the system's expected Young's angle is subsequently computed with Eq.~\ref{eqn:YoungsAngle}.

\begin{figure}
\includegraphics{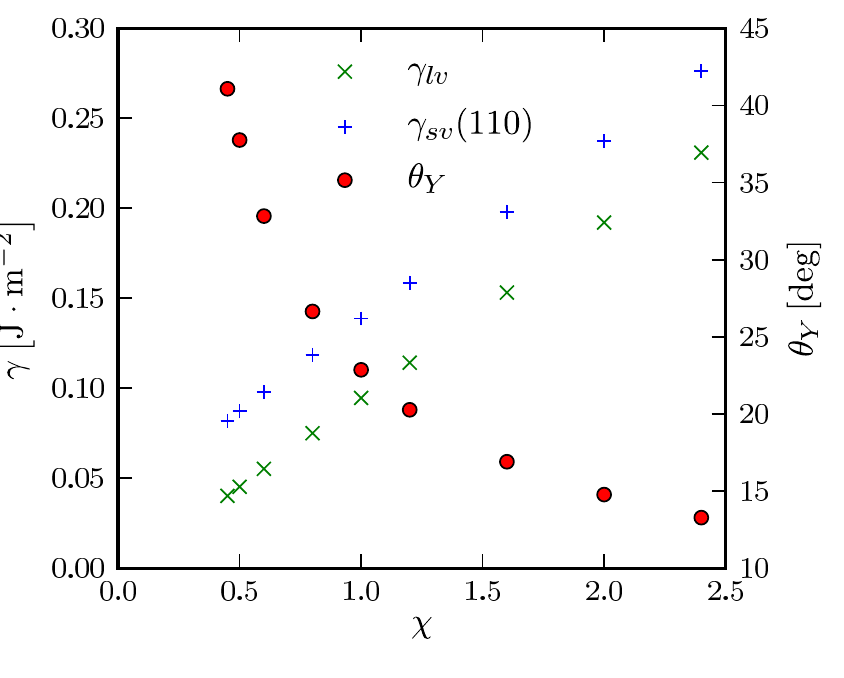}
\caption{(Color online) Surface energies for $\left(110\right)$ solid-vapor $\gamma_{sv}$ and liquid-vapor $\gamma_{lv}$ interfaces, and the resulting Young's angle $\theta_Y$ as a function of $\chi$.}  
\label{fig:ThetaY}
\end{figure}

Figure.~\ref{fig:ThetaY} shows the measured surface energies and confirms they are both linearly proportional to $\sqrt{\kappa W}$.  Additionally, this figure suggests that the quantity $\theta_\text{Y}$ can effectively be tuned to a desired value by modifying $\sqrt{\kappa W}$ at constant $\sqrt{\kappa / W}$.  In practice, obtaining equilibrium trijunctions in numerical simulations with periodic boundary conditions is challenging due to the Gibbs-Thompson effect.  We have observed trijuncitons out of equilibrium, and intend to address their motion in future work.  In the next two sections, we consider the liquid-vapor and solid-vapor interfaces in greater detail.

\subsection{Liquid-Vapor Interface Structure}
\label{sec:liquid-vapor}
PFC simulations of equilibrium liquid-vapor interfaces, including the system displayed in Fig.~\ref{fig:threePhaseSystem-trace}b.), indicate that there is significant structure to the $\rho$ field near the liquid-vapor interface.  Specifically, the PFC model produces liquid-vapor interfaces with oscillations in the liquid density that decay in amplitude with increasing depth into the liquid.  This type of interface structure has been observed in both simulations and experiments for other liquid-metal vapor interfaces~\cite{Penfold:2001,Develyn:1981,Regan:1997,Zhao:1998,Gonzalez:2007}.  According to D'Evelyn and Rice~\cite{Develyn:1981}, the rapid decrease in conduction electron density from the liquid to the vapor induces an abrupt change in the pair interaction potential which induces atomic stacking on the liquid side of the interface.  In the present PFC model, the modulation of $C_2$ by the interpolating function $g\left( \eta \right)$ produces a similar effect, and the PFC interfaces indeed exhibit density oscillations.  In this section, we consider only the liquid and vapor phases and show that the PFC model can quantitatively reproduce some features of liquid-vapor interface structure as determined by both experiments and more sophisticated models.

We employ a correlation function with a single peak whose position $k_1$, height $\hat{C}\left( k_1 \right)$, and second derivative at the peak $\hat{C}''\left( k_1 \right)$ are set to match measurements of $S\left( k \right)$ by neutron and x-ray diffraction for liquid Gallium\cite{Narten:1972}.  Additionally, we choose the parameters $\kappa$ and $W$ to match both the experimentally determined surface energy\cite{Hardy:1985}, and approximate the density profile width\cite{Regan:1997}.  First, we compute $\hat{C}\left(k\right)$ from the experimental $S\left( k \right)$ using Eq.~\ref{eqn:Ck_Sk_relation}, and then fit this with a parabola in the region of the first peak $k_1 = 2.5\text{\AA}^{-1}$\cite{Narten:1972}.  The values $\hat{C}_2\left(k_1\right)$ and $\hat{C}''_2\left(k_1\right)$ are determined from the fit coefficients, and Eqs.~\ref{eqn:tempScale}-\ref{eqn:alphaScale} are employed to set the parameters $T_0$ and $\alpha_1$ to match the experimental peak properties.  Table~\ref{tbl:GaProps} summarizes the values extracted from the experimental $S\left( k \right)$ as well as the resulting PFC parameters.  Finally, $\rho_0$ is chosen to be the experimentally determined liquid density.

The quantities $\psi_v^0=-0.99$ and $\Delta=0.0$ are set to ensure that equilibrium liquid and vapor densities are close to $\rho_0$ and 0 respectively.  The bulk modulus $K_i$ of the homogeneous phase $i$ is
\begin{equation}
	K_i \propto { \bar{\rho}_i \, }^2 \left( \frac{\partial^2 f}{\partial \psi^2}\right)\bigg|_{\bar{\psi}_i}.
\end{equation}
For our system where $\bar{\rho}_l \approx \rho_0$ and $\bar{\rho}_v \approx \rho_0 \left(1+\psi_v^0 \right)$, 
\begin{equation}
\frac{K_v}{K_l} \approx  \left( 1 + \psi_v^0 \right)^2 b / \left(1 - A_0 \right)
\end{equation}
With $\psi_v^0 =-0.99$, $K_v / K_l \approx 10^{-4} $ for $b=1$ and $A_0=0$.  This bulk modulus ratio is reasonable for common liquids and vapors near atmospheric pressure.  Finally, the parameters $\nu$ and $\xi$ have only a weak influence on the liquid-vapor surface properties, and thus in this section they are both set to 1 for simplicity.

\begin{table}
\caption{Model parameters used to simulate Gallium liquid-vapor interface at $T_\text{ref}=293$K.  $k_1$, $\rho_0$, $\hat{C}_2\left( k_1 \right)$, and $\hat{C}_2''\left( k_1 \right)$ are taken from data in Ref.~\citenum{Narten:1972}, and the value of $\gamma_{lv}$ is reported in Ref.~\citenum{Hardy:1985}.}
\label{tbl:GaProps}
\begin{ruledtabular}
\begin{tabular}{l d c}
	Quantity 				& \text{Value}		 & \text{Unit} 		\\ \hline
	$k_1$					& 2.50 \times 10^{10}& $\text{m}^{-1}$	\\
	$\rho_0$				& 5.28 \times 10^{28}& $\text{m}^{-3}$	\\
	$\hat{C}_2\left( k_1 \right)$	& 0.60		 & - 				\\
	$\hat{C}_2''\left( k_1 \right)$	& -10.84 \times 10^{-20}	& $\text{m}^2$ \\
	$\gamma_{lv}$			& 0.714				& $\text{J} \cdot \text{m}^{-2} $  \\
	$\alpha_1$				& 2.35 \times 10^{9}& $\text{m}^{-1}$	\\
	$T_0$					& 410.				& $\text{K}$ 		\\
	$T$						& 293.				& $\text{K}$		\\
	$\psi_v^0$				& -0.99				& -					\\	
	$\Delta$				& 0.0& -					\\	
	$\nu, \xi, b$			& 1.0				& -					\\
	$W$						& 560				& -					\\
	$\kappa$				& 3.5\times 10^{-19}&$\text{m}^2$
\end{tabular}
\end{ruledtabular}
\end{table}

Regan et~al.\cite{Regan:1995,Regan:1997} determined the density profile of a liquid Gallium-vapor interface using X-ray reflectivity data measured at room temperature.  Their reflectivity results were well described by the empirical density profile model
\begin{align}
	\frac{ \rho \left(  z' \right) - \bar{\rho}_v } { \bar{\rho}_l - \bar{\rho}_v } &= \frac{1}{2}\left[ 1 + \text{erf} \left( \left(z' - \Delta z\right)/\delta \right) \right] \notag \\
	&+ H\left( z'\right) A \sin \left( 2 \pi z' / \lambda \right) \exp \left( -z'/\zeta \right),
	\label{eqn:LV-model}
\end{align}
where $z' = z- z_0$, $z$ is distance normal to the interface with positive $z$ into the liquid, $z_0$ is the interface location, $\Delta z$ is an offset, $\delta$ is a measure of the interface thickness, $A$, $\lambda$, and $\zeta$ are the amplitude, wave length, and decay length of density oscillations on the liquid side of the interface, and $H\left( z' \right)$ is a step function centered at $z' = 0$.  We fit Eq.~\ref{eqn:LV-model} to the numerically determined PFC density profile and Table~\ref{tbl:intFitParams} summarizes the fit parameters for the PFC model as well as experimental results from Ref.~\citenum{Regan:1997}.  We find that the parameters for the liquid-vapor surface structure produced by the PFC model at 293 K are in quantitative agreement with experimental results for Gallium, with the exception of the oscillation amplitude which is roughly $4\times$ smaller in the PFC result.  The oscillation decay and wavelength are also in general agreement with more complex simulation techniques including the self-consistent quantum Monte Carlo simulations of Zhao et al.\cite{Zhao:1998}, and orbital-free ab initio molecular dynamics simulations of Gonz\'{a}lez et al.~\cite{Gonzalez:2007}.

\begin{figure}
\includegraphics{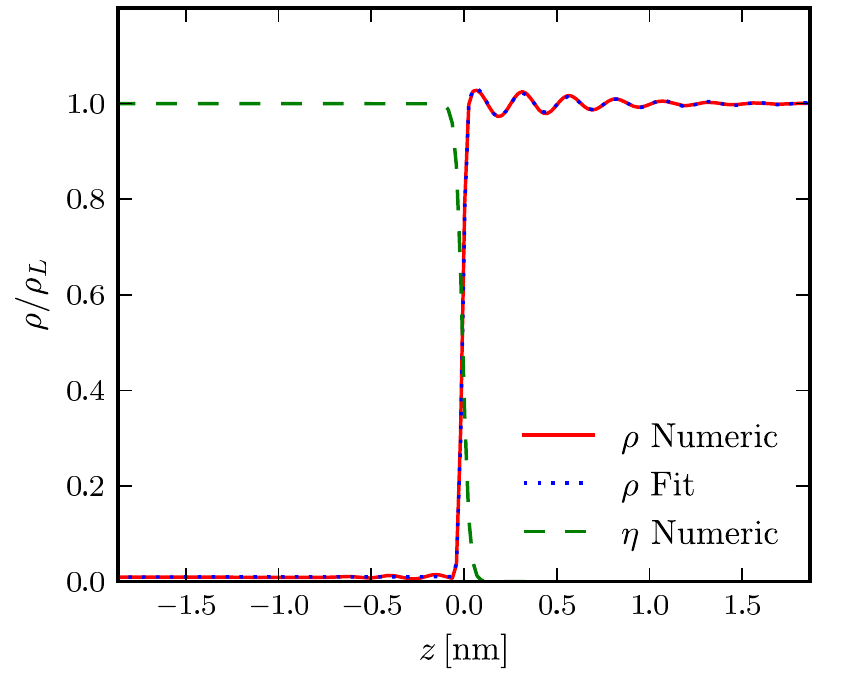}
\caption{(Color online) Numerically relaxed density $\rho$ (solid line) and order parameter $\eta$ (dashed line) for a Gallium liquid-vapor interface at $T = 293$~K.  The fit (dotted) exhibits oscillation wavelength and amplitude decay length in agreement with experiments for Gallium~\cite{Regan:1997} as listed in Table~\ref{tbl:intFitParams} .}
\label{fig:densityFit}
\end{figure}

\begin{table}
\caption{Fitting parameters for Eq.~\ref{eqn:LV-model} for experimental and PFC profiles.  All experimental Gallium data from Ref.~\citenum{Regan:1997}.  The numerical surface energy is $0.715 \,\,\text{J} \cdot \text{m}^{-2}$, and Ref.~\citenum{Hardy:1985} indicates an experimental value $0.710\,\,\text{J} \cdot \text{m}^{-2}$.  Uncertainties indicate standard errors of the estimated parameters returned by an orthogonal distance regression routine\cite{Boggs:1992}.}
\label{tbl:intFitParams}
\begin{ruledtabular}
\begin{tabular}{l  l  l  }
	Parameter 	& Ga Ref.~\citenum{Regan:1997}	& PFC  \\ \hline
	$\lambda \left[ \text{nm} \right]$	& $0.256 \pm 0.001$		& $0.252	\pm 3.3 \times 10^{-4}$	\\
	$\zeta   \left[ \text{nm} \right]$	& $0.580 \pm 0.04$		& $0.622	\pm 0.011$				\\
	$\delta  \left[ \text{nm} \right]$	& $0.050 \pm 0.004$		& $0.024 \pm 4.6 \times 10^{-4}$	\\
	$A \left[ \rho / \rho_0 \right]$	& $0.20 \pm 0.02$		& $0.037 \pm 5.2 \times 10 ^{-4}$   \\
\end{tabular}
\end{ruledtabular}
\end{table}

\subsection{Solid-Vapor Interfaces}
\label{sec:solid-vapor}
\begin{figure}
\includegraphics{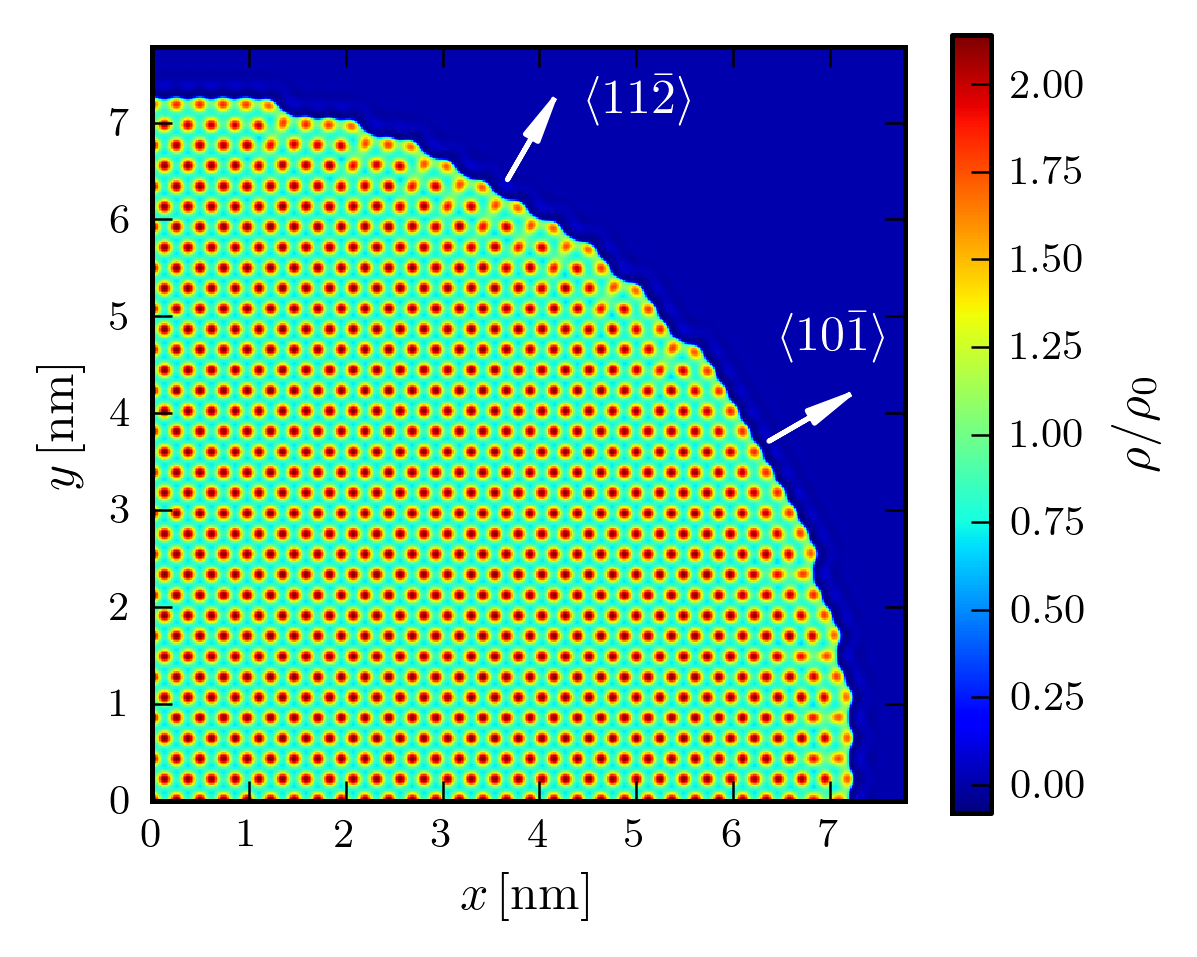}
\caption{(Color online) Interface between solid particle and surrounding vapor phase.  Facet interface normals of type $\langle 1 0 \bar{1} \rangle$ and $\langle 1 1 \bar{2} \rangle$ are indicated by arrows, and steps are clearly visible.}
\label{fig:hexParticleCloseup}
\end{figure}

Next, we describe several aspects of solid-vapor interface structure for PFC simulations below $T_{tr}$.  For simplicity, we consider two-dimensional systems and employ only one peak in $\hat{C}_2$.  These choices favor periodic states with hexagonal symmetry when cooled below $T_{tr}$ as shown in Fig.~\ref{fig:hexParticleCloseup}.

\begin{table}
\caption{Parameters for the 2D hexagonal system.}
\label{tbl:HexParams}
\begin{ruledtabular}
\begin{tabular}{l d c}
	Quantity 				& \text{Value}		 & \text{Unit}      \\ \hline
	$k_1$					& 2.985\times 10^{10}& $\text{m}^{-1}$  \\
	$\rho_0$				& 6.57 \times 10^{28}& $\text{m}^{-3}$  \\
	$\alpha_0$				& 2 \alpha_1         & $\text{m}^{-1}$  \\
	$\alpha_1$				& 5.07 \times 10^{9} & $\text{m}^{-1}$  \\
	$T_0$					& 1820.				 & $\text{K}$       \\
	$T$						& 1000.				 & $\text{K}$       \\
	$\psi_v^0$				& -0.99				 & -				\\	
	$\Delta$				& -0.115             & -				\\	
	$\nu, \xi, b$			& 1.0				 & -				\\
	$W$						& 1.0				 & -				\\
	$\kappa$				& 1\times 10^{-21}   &$\text{m}^2$
\end{tabular}
\end{ruledtabular}
\end{table}

As described briefly in Sec.~\ref{sec:equib}, the solid-vapor interface is significantly more complicated than the isotropic liquid-vapor interface due to the anisotropy of the solid phase.  As with the liquid-vapor interfaces, there is a sharp decrease in the influence of $C_2$ across the solid-vapor interface and the periodic nature of the crystalline density field decays to the homogeneous vapor density through a width of approximately $a_0$.  Figure~\ref{fig:hexParticleCloseup} shows a portion of the interface between a crystalline particle surrounded by vapor.  This interface consists of two distinct types of facet truncated by steps, features characteristic of anisotropic solid-vapor interfaces.  Using the 2D hexagonal basis vectors shown in Fig.~\ref{fig:SV_interface_schematic}a.), the facets in Fig.~\ref{fig:hexParticleCloseup} have interface normals along $\langle 1 0 \bar{1} \rangle$ and $\langle 1 1 \bar{2} \rangle$ type directions.  In the next section, we describe the change in the excess surface free energy with respect to changes in step spacing in order to measure the excess step free energy. Then, we briefly discuss the elastic strain field in the crystal below such a stepped surface.  All simulations in this section are for 2D hexagonal crystals and are carried out with parameters from Table~\ref{tbl:HexParams} unless otherwise noted.

\subsubsection{Solid-vapor step energy} 
\label{sec:SV-gamma_anisotropy}
\begin{figure}
	\includegraphics{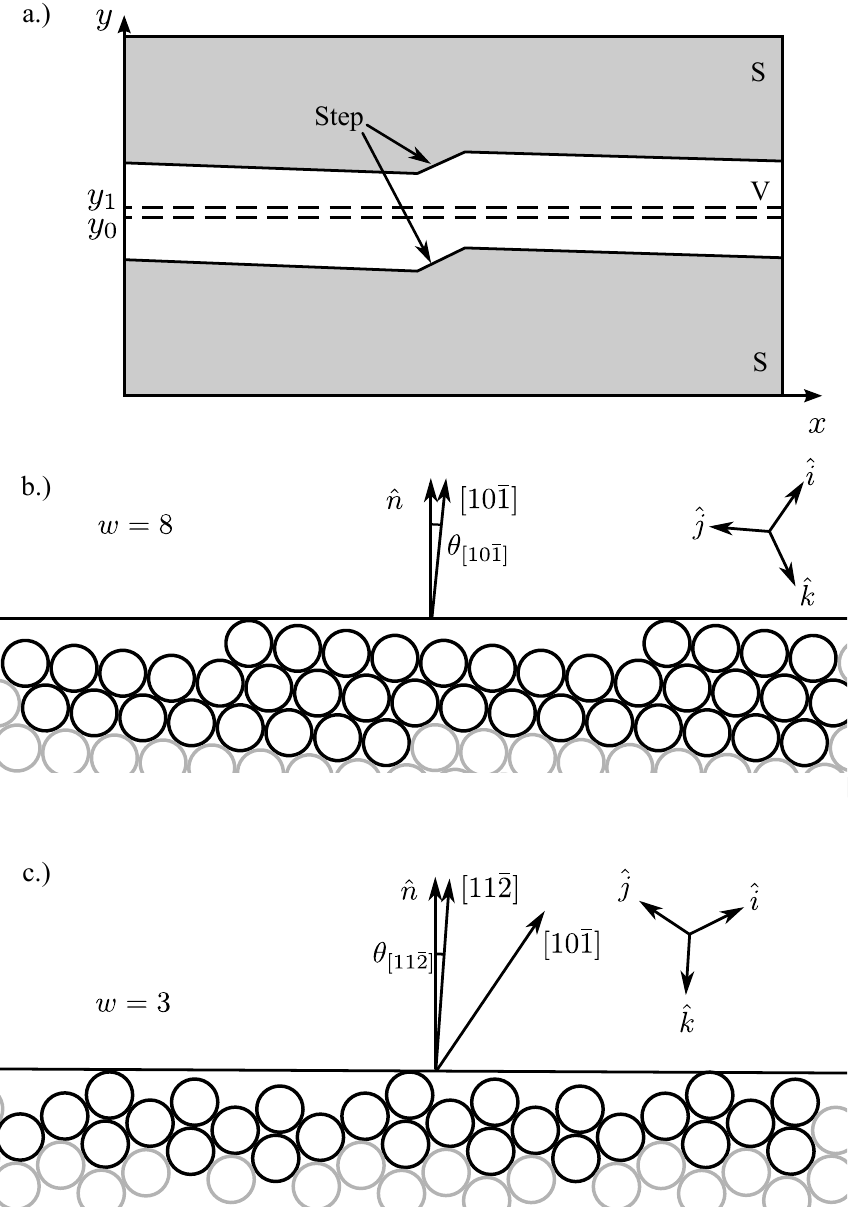}
	\caption{a.) Schematic of the simulation domain employed to test solid-vapor interface properties and step flow growth for 2D hexagonal solids.  Crystalline phase is shaded and labeled $S$ and vapor $V$.  The domain has dimensions $l_x \times l_y$, all boundaries are periodic, and for simulations where solid growth is induced, the matter source is located at $y_0 < y < y_1$, and the interface normal $\hat{n}$ is along the $y$ axis.	 b.)~Geometry of a stepped surface with facet normals of $ \left[ 1 \, 0 \, \bar{1} \right] $ and width $w=8$.  The horizontal solid line is the plane of the macroscopic interface.  c.)~Geometry of a stepped surface with facet normals $ \left[ 1 \, 1 \, \bar{2} \right] $ and width $w=3$.  Hexagonal basis vectors $\hat{i}, \hat{j}, \hat{k}$ are included in b.) and c.). }
	\label{fig:SV_interface_schematic}
\end{figure}

The solid-vapor interface for the particle in Fig.~\ref{fig:hexParticleCloseup} exhibits two types of crystallographic facets truncated by steps, reflecting the anisotropic nature of the solid-vapor surface.  Step energy is an important factor in the growth of faceted crystals, and recent MD simulations have shown that these energies can be quantified by measuring changes in coexistence temperatures and island radius\cite{Frolov:2012}.  In this work, we quantify the excess energy of the step by varying the spacing of a periodic array of steps and measuring the change in surface energy.  First, we prepared initial conditions with a slab of solid and vapor as shown schematically in Fig.~\ref{fig:SV_interface_schematic}a).  The domain dimensions are selected to accommodate a periodic array of steps as described below.  A sharp cutoff in density between the solid and vapor is allowed to relax, forming a step, and the surface energy is measured numerically as before.

For interfaces with steps on facets with normals of type $\langle 1 0 \bar{1} \rangle$ as in Fig.~\ref{fig:SV_interface_schematic}b.), we define $w$ as the integer number of peaks on the facet with $\leq 4$ nearest neighbors.  The angle between the interface normal and the facet normal depends on the facet type and $w$.  For interfaces with $\left[ 1 \, 0 \, \bar{1} \right]$ facets, this angle is given by the geometric relationship
\begin{equation}
	\sin \theta_{\left[ 1 0 \bar{1} \right]} = \frac{\sqrt{3}}{2} \left( w\left(w + 1\right) + 1 \right)^{-1/2},
	\label{eqn:theta101}
\end{equation}
which is used to select the domain dimensions and orientation of the crystal in the initial condition.  For the interfaces with facet normal of type  $\langle 1 1 \bar{2} \rangle$ as in Fig.~\ref{fig:SV_interface_schematic}c.), we define $w$ as the number of peaks on the terrace having 3 nearest neighbors, and then
\begin{equation}
	\sin \theta_{\left[ 1 1 \bar{2} \right]} = \frac{1}{2} \left(3 w\left( w + 1 \right) +1\right)^{-1/2} .
	\label{eqn:theta112}
\end{equation}
Note that in both cases, the macroscopic plane of the interface is parallel to one edge of the simulation domain as shown by the vector $\hat{n}$ in Fig.~\ref{fig:SV_interface_schematic}b-c.

For interfaces with equally spaced steps, and under the assumption that the step energy density $\beta_{\hat{n}}$ is not a function of spacing but does depend on the facet orientation $\hat{n}$, the simplest model of the excess free energy of the surface is~\cite{Srolovitz:1991}
\begin{equation}
	\gamma \left( \theta \right) = \gamma_{\hat{n}} + \frac{\beta_{\hat{n}}}{h_{\hat{n}}} \, \left| \theta_{\hat{n}} \right|,
	\label{eqn:gammaVsTheta_small}
\end{equation}
where $\sin \theta_{\hat{n}} \approx \theta_{\hat{n}}$ can be computed from $w$ using either Eq.~\ref{eqn:theta101} or~\ref{eqn:theta112}, $\gamma_{\hat{n}}$ is the excess free energy of the facet with infinite step spacing, $\beta_{\hat{n}}$ is the excess free energy per step, $h_{\hat{n}}$ is the step height, and $\hat{n}$ is the facet normal, either $\left[ 1 0 \bar{1} \right]$ or $\left[ 1 1 \bar{2} \right]$ for our measurements.

\begin{figure}
\includegraphics{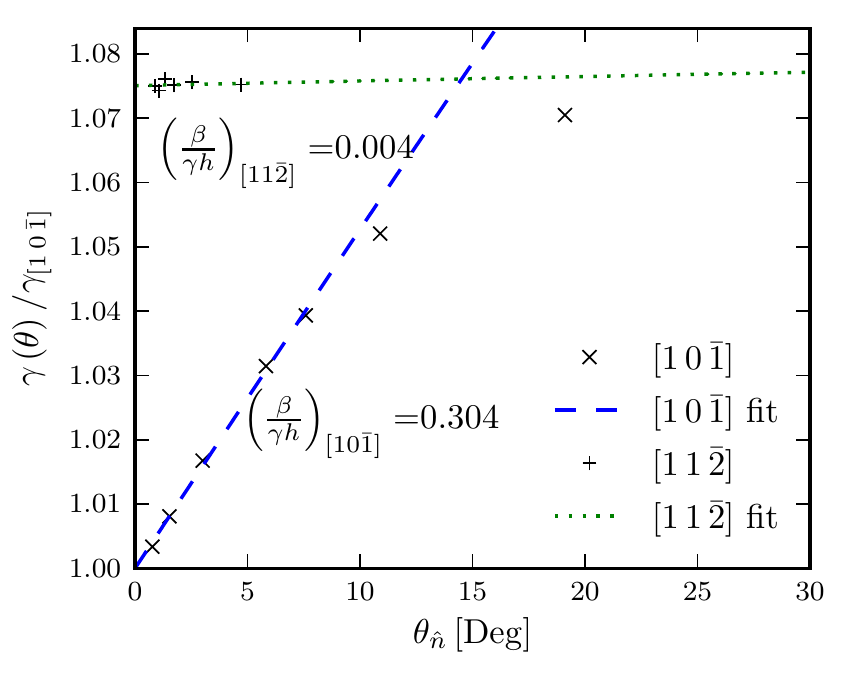} 
\caption{(Color online) Surface energy $\gamma$ as a function of the angle between the interface and facet normals for interfaces with facet normals $ \left[ 1 \, 0 \, \bar{1} \right] $ $\left( \times \right)$ and $\left[ 1 \, 1 \, \bar{2} \right]$ $\left( + \right)$.  Fits of Eq.~\ref{eqn:gammaVsTheta_small} to data points for $\left[ 1 \, 0 \, \bar{1} \right]$ (dashed) and $\left[ 1 \, 1 \, \bar{2} \right]$(dotted) facets are included.}
\label{fig:hex_gamma_vs_theta}
\end{figure}

Figure~\ref{fig:hex_gamma_vs_theta} shows the measured excess free energy $\gamma\left( \theta \right)$ normalized by the $\left[ 1 \, 0 \, \bar{1} \right]$ facet energy for stepped interfaces with facets of both $\left[ 1 \, 0 \, \bar{1} \right]$ and $\left[ 1 \, 1 \, \bar{2} \right]$ type as a function of the angle between the interface and facet normals.  Values of $\beta_{\hat{n}}$ and $\gamma_{\hat{n}}$ are determined by fitting Eq.~\ref{eqn:gammaVsTheta_small} to the experimental data for both facet types, and the dimensionless quantity $ \left( \beta / \left( \gamma h\right) \right)_{\hat{n}}$ is then computed from the fitting parameters.  The measured values for steps on $\left[ 1 \, 0 \, \bar{1} \right]$ and $\left[ 1 \, 1 \, \bar{2} \right]$ facets are 0.304 and 0.004 respectively.  These values are in the same range as that reported for cubic transition metals using first principles and cluster expansion methods\cite{Vitos:1999}.   The more closely packed $\left[1 0 \bar{1} \right]$ plane has a lower surface energy, and a larger value of $\beta / \left( \gamma h \right)$ as expected based on the number of missing neighbors.

\subsubsection{Elastic field}
We measure the elastic strain field below stepped surfaces predicted by the PFC model by comparing the positions of the local maxima of the $\rho$ field to the expected positions for a bulk crystal with no interfaces.  Appendix~\ref{sec:peakFinding} contains a description of the procedure used to determine density peak coordinates.  We then use the peak coordinates to determine  $\Delta d_{n,n+1}$,  the change in the spacing  between the $n^\text{th}$ and $n+1^\text{th}$ crystallographic planes parallel to the macroscopic interface as described by Srolovitz and Hirth~\cite{Srolovitz:1991}.  An expansion (contraction) of the plane spacing compared to the bulk value is indicated by $\Delta d_{n,n+1} > 0 \, \left( < 0 \right)$.  We consider 2D hexagonal crystals and use the same parameters as described in Table~\ref{tbl:HexParams}.  Finally, note that the plane spacing can be computed directly from the $y$ coordinate for the relevant peaks, as shown in Fig.~\ref{fig:SV_interface_schematic}a).

\begin{figure}
\includegraphics{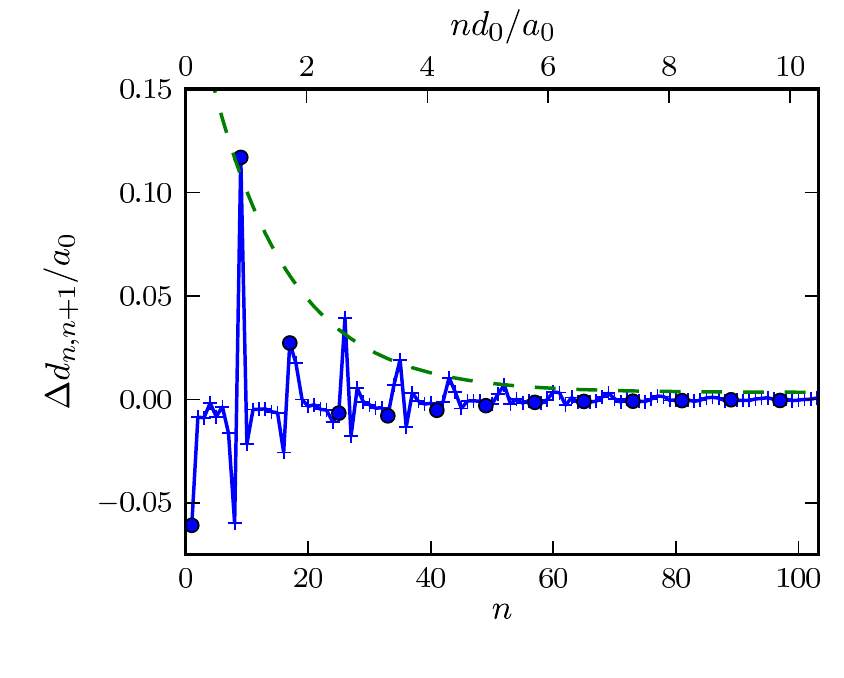} 
\caption{(Color online) Plane spacing change $\Delta d_{n,n+1}$ below a step on a solid-vapor interface of a 2D hexagonal crystal as a function of layer index $n$ ($n=1$ is the plane including the step).  The system has $w=8$ peaks separating the steps as shown in Fig.~\ref{fig:SV_interface_schematic}a.), and spacing change is normalized by the nearest neighbor spacing $a_0$.  Solid filled circles indicate $n = 8\,m + 1$ for $m=0,1,2,...$.  The second horizontal axis gives the depth $n d_0$ normalized by the nearest neighbor spacing $a_0$, where $d_0$ is the ideal plane spacing.  The dashed curve is an approximate envelope of the local maxima that indicates an exponential decay with increasing depth into the solid.}
\label{fig:PlaneSpacing}
\end{figure}

Figure~\ref{fig:PlaneSpacing} shows $\Delta d_{n,n+1}$ measured for equilibrated surfaces.  The facet normals are $\left[1 \, 0 \, \bar{1} \right]$, and $w=8$.  The spacing between the first two planes, $\Delta d_{1,2}$, exhibits a contraction of 5~\% of $a_0$, and the value of $\Delta d_{8,9}$  (immediately below the step) shows an expansion of 12~\% of $a_0$.  Subsequent layers show periodic expansions and contractions, with the magnitude of $\Delta d_{n,n+1}$ decreasing with increasing depth into the solid.  Similar trends are observed for the step spacings $w=4,16,32$ (not shown), and the amplitudes $\Delta d_{n,n+1}$ for $n=1$ and $n=w$ are approximately the same for all values of $w$ tested.  The period of the oscillation is approximately equal to the inter-atomic spacing and is independent of surface orientation.  These behaviors are both consistent with the results of Chen, Voter, and Srolovitz~\cite{Chen:1986,Chen:1989}.  The approximate dashed envelope in Fig.~\ref{fig:PlaneSpacing} indicates an exponential decay in the oscillation amplitude and is also consistent with Chen et~al.~for various surfaces of aluminum and nickel\cite{Chen:1986,Chen:1989} and experimental values for aluminum and copper (110) surfaces~\cite{Adams:1982,Nielsen:1982}.   

While the decay and wavelength are in agreement with other simulations and experiments, the absolute value $\Delta d_{1,2}$ measured in our simulations is an order of magnitude greater than these results.  The elastic constants of the crystal phase are proportional to ${\alpha_1}^{-2}$~\cite{Greenwood:2011a}.  In the present work, $\alpha_1$ was chosen such that the strains were large enough to be easily resolvable on grids with $a_0/\Delta x \approx 16$, making computations numerically tractable, rather than to match any particular material constants.  Finally, we note that the slow decay of strain into the bulk indicates that a thick slab of solid is necessary to reduce the effect of a finite system size on the strain measurements.  The slab half-thickness for the simulation in Fig.~\ref{fig:PlaneSpacing} is approximately $60 a_0$, but larger domains might be required depending on simulation parameters.

\subsubsection{Step-Flow Growth}
In previous sections we have focused on the equilibrium behaviors of the model.  In this section, we test the dynamical behavior of the model with simulations of step flow growth of a two-dimensional solid in contact with a supersaturated vapor at temperatures below the triple point.  The system consists of slabs of vapor and solid with periodic boundary conditions in both directions similar to the systems described in Fig.~\ref{fig:SV_interface_schematic}a.).  As the domain is periodic, the interface effectively has an infinite number of equally spaced steps, and nucleation of new steps is not necessary for continued growth.  We induce solid growth by introducing matter into the domain via the addition of a source term $S_\psi\left( \vec{r} \right)$ to the right hand side of Eq.~\ref{eqn:psiEvo}.  Specifically, 
\begin{equation}
S_\psi =
	\begin{cases}
		s_0, & y_0 < y < y_1 \\
		0, & \text{otherwise,}
	\end{cases}
\end{equation}
where $s_0$ is the source strength, $y$ is the coordinate normal to the interface, and $y_0$ and $y_1$ are $y$ values chosen to locate the source in a layer parallel to the solid-vapor interface  in the center of the vapor slab.  This source causes a linear increase in $\bar{\rho}$ with time, biasing the system toward higher solid volume fraction and causing the solid phase to grow.  Note that as the domain is periodic, the source feeds the growth of the solid slab on both faces.

We define the mean chemical potential in the vapor phase $\bar{\mu}_v$
\begin{equation}
	\bar{\mu}_v = \sum_i g\left(\eta_i\right) \mu_i \bigg/ \sum_i g\left(\eta_i\right),
\end{equation}
where the sum is over all grid points contained in the region $0 < z < l_y / 2$ in Fig.~\ref{fig:SV_interface_schematic}a.).  The normalized mean driving force for solid growth $\Delta \tilde{\mu}$ is defined
\begin{equation}
	\Delta \tilde{\mu} = \left( \bar{\mu}_v - \mu_0\right) / \mu_0,
	\label{eqn:supersat}
\end{equation}
where $\mu_0$ is the equilibrium chemical potential of a system with solid and vapor in coexistence.

\begin{figure}
	\includegraphics{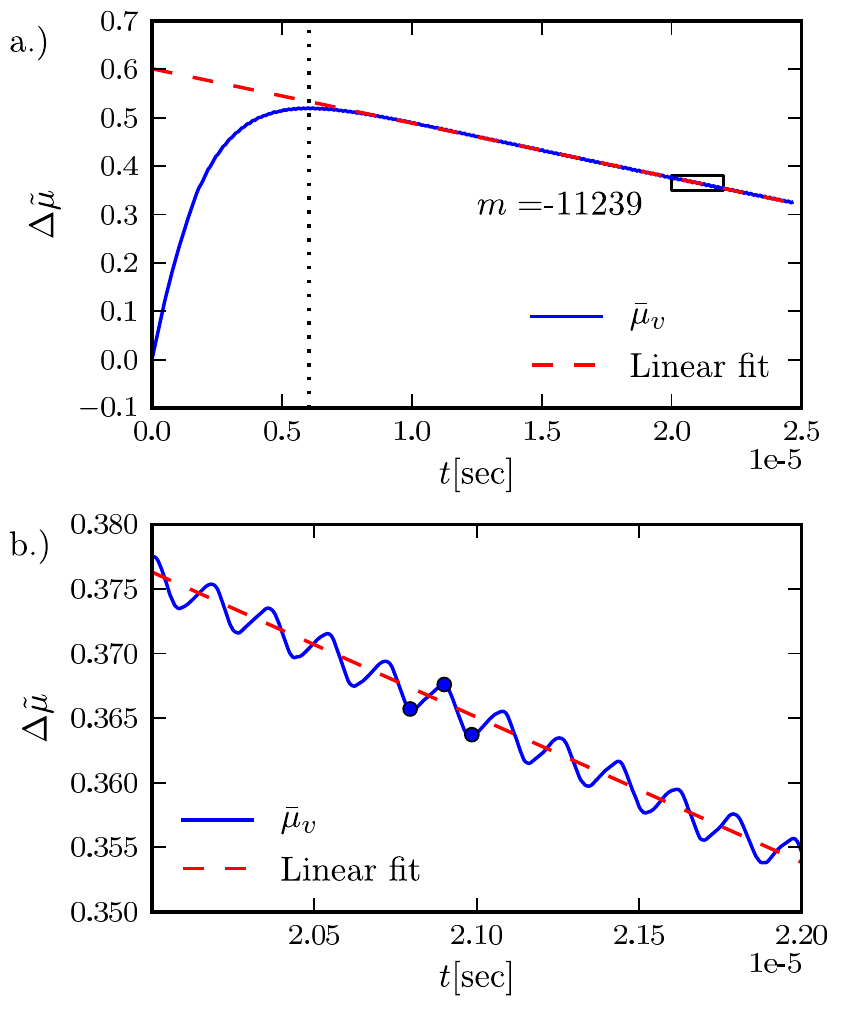}
	\caption{(Color online) a.) Normalized mean driving force $\Delta \tilde{\mu}$ from Eq.~\ref{eqn:supersat} during step flow growth with a linear fit to the late stages of growth.  There is an initial transient before $t^*= {l_y}^2 / \left( 4 M_\rho \right) \approx 6 \times 10^{-6} $~s marked by a vertical dotted line.  b.) Inset from the rectangle shown in a.) showing oscillation of the driving force for growth with period $t = 1.8\times10^{-7}$s.  Each period corresponds to the addition of one density peak to the solid.  Solid circles correspond to the times shown on Figs.~\ref{fig:stepFlowChemPotMap}a.)-c.).}
	\label{fig:stepFlow-ChemPotTime}
\end{figure}

This system is first allowed to come to equilibrium with $s_0 = 0$, and the equilibrium chemical potential $\mu_0$ is computed numerically.  The source is then turned on, and the vapor phase begins to supersaturate.  Figure~\ref{fig:stepFlow-ChemPotTime} displays $\Delta \tilde{\mu}$ as a function of time with the source turning on at $t=0$.  After an initial transient period $t \lesssim 6 \times 10^{-6}$ s, $\Delta \tilde{\mu}$ exhibits oscillatory behavior with a period of $t=1.8 \times 10^{-7}$ s.  Each period of the oscillation corresponds to the addition of one density peak to the crystal, with the step advancing a distance $a_0$ tangent to the interface.  The measured period for peak addition and the mean interface velocity over the full simulation are both in good agreement with estimates based on the source strength and assumptions of steady state growth.  The oscillating behavior of the driving force is similar to that observed by Tegze et al. for PFC simulations of layer by layer growth of a crystal into a liquid phase\cite{Tegze:2009}, although in that case oscillations correspond to entire layers being added to the crystal rather than single peaks.

Figure~\ref{fig:stepFlow-ChemPotTime}~a.) also indicates that there is a slow overall decrease in $\Delta \tilde{\mu}$ after the initial transient period.  This behavior is expected due to the decreasing distance between the growth interface and the source.  The average flux of material to the growth interface is directly determined by $s_0$ and is therefore constant.  Because the mean chemical potential gradient in the vapor is proportional to the mass flux, it is also constant on time scales longer than period for peak addition.  Assuming the chemical potential of the solid does not change during growth, $\Delta \tilde{\mu}$ must decrease accordingly.  The slope of the dashed fit line in Fig.~\ref{fig:stepFlow-ChemPotTime}~a.) is in resonably good agreement with an estimate computed based on the above assumptions and the value of $s_0$.

\begin{figure}
	\includegraphics{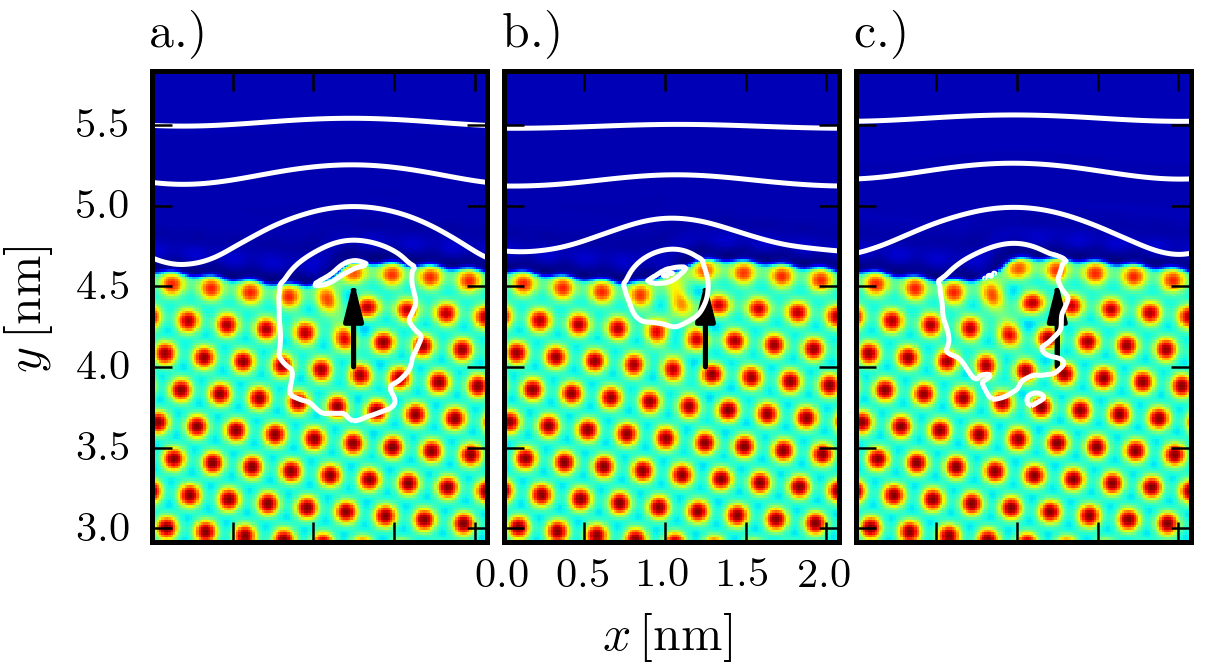}
	\caption{(Color online) Density field at times marked with circles in Fig.~\ref{fig:stepFlow-ChemPotTime}b). Chemical potential contours are shown as white lines and the arrow indicates the location of the step at the time shown in a.).}
	\label{fig:stepFlowChemPotMap}
\end{figure}

Figure~\ref{fig:stepFlowChemPotMap} displays a portion of the $\rho$ field near the growth interface, as well as contours of $\mu$ at the three instances labeled in Fig.~\ref{fig:stepFlow-ChemPotTime}b.).  The $\mu$ difference between contours is uniform across all three plots and therefore the physical spacing between contours indicates relative changes in the $\mu$ gradient and therefore $\rho$ flux.  Figure~\ref{fig:stepFlowChemPotMap}a.) shows that at $t=2.0795\times10^{-5}$~s, when the system has the lowest driving force just after the addition of a new density peak, the $\mu$ field is relatively flat within the crystal, but exhibits a gradient in the vapor that is roughly normal to the crystal-vapor interface.  At $t=2.09\times10^{-5}$~s, the driving force has increased to its highest level, and the $\mu$ contours in Fig.~\ref{fig:stepFlowChemPotMap}b.) indicate strong flux toward the step, which is a local minimum in $\mu$, as the excess matter in the vapor phase flows into a new density peak at the former step location.  Finally, addition of the new density peak is complete by $t=2.0985\times10^{-5}$~s, and the system returns to state similar to that in Fig.~\ref{fig:stepFlowChemPotMap}a.), and the process begins again.  The $\mu$ value within the crystal is nearly uniform throughout this process, consistent with assumptions described in the previous paragraph.

\section{Conclusions}
\label{sec:Conclusions}
This work extends existing PFC models of liquid and solid systems to include a vapor phase.  This is accomplished by constructing a free energy functional that modulates the strength of the two body correlation function, where the amplitude of the correlation term is strong in the condensed phases and zero in the vapor phase.  This extension enables the study of processes which involve crystal-vapor and liquid-vapor interfaces, in addition to systems with crystal-crystal and crystal-liquid phase transformations addressed by classic PFC models.  This feature allows the model to tackle questions of important materials processing scenarios such CVD and VLS growth over diffusive time scales.

The theoretical and numerically computed phase diagrams display characteristic behaviors for a pure material near its triple point.  Other features of the phase diagram, such as the slopes of the liquid-vapor phase boundaries, could be modified by making the appropriate model parameters temperature dependent for physical problems where these features are essential.  Also, the rapid decay of the two-body correlation term gives rise to curvature independent interface effects that cause shifts in the phase boundaries between vapor and condensed phases.  The numerically measured shifts agree quantitatively with theoretical predictions.

Our simulations exhibit density oscillations on the liquid side of equilibrium planar liquid-vapor interfaces.  These profiles are in quantitative agreement with theoretical and measured profiles for liquid metals near their triple point\cite{Regan:1995,Zhao:1998,Gonzalez:2007}.  While the model free energy is phenomenological, this qualitative agreement with experiment indicates that the modulation of $C_2$ is plausible for metals where there is a sharp transition from conductor to insulator across the liquid-vapor interface.

The PFC model also produces strongly anisotropic solid-vapor interfaces with terraces of low index facets truncated by steps, and the relative magnitude of the step energy with respect to the surface energy lies in the range predicted for transition metals~\cite{Vitos:1999}.  Also, the elastic strain field in the vicinity of the steps agrees qualitatively with data from experiments and simulations~\cite{Adams:1982,Nielsen:1982,Chen:1986,Chen:1989}.  Finally, in the presence of a mass source in the vapor phase these crystals undergo step-flow growth, an important process in CVD.

The breadth of physical phenomena that the model reproduces, combined with the diffusive time scales over which it operates, make this an excellent tool for investigating many technologically relevant processes.  We anticipate that an extension of the present model to include alloys will enable the study of more diverse three-phase systems~\cite{Elder:2007,Elder:2010b,Elder:2011,Greenwood:2011b}.  This could include the study of phenomena such as trijunction motion over faceted solids, the interaction of tri-junctions with crystal defects like grain boundaries or dislocations, and, ultimately, technologically important processes such as VLS nanowire growth in a binary alloy system.

We have used the correlation function based PFC model of Greenwood~et~al.~\cite{Greenwood:2010,Greenwood:2011a} for the condensed phases rather than the classic free energy~\cite{Elder:2002} in order to have more control over crystal symmetry.  We have limited ourselves to 2D hexagonal and 3D BCC crystals with only one peak in the correlation function in the present work.  It is to be seen how the addition of more peaks affects quantities such as the solid-vapor step energy for both these simple crystals, as well as for more complex structures.

Finally, we also anticipate that the vapor phase will be useful for the investigation of several phenomena that have been previously been studied with a liquid as a stand in for the vapor phase.  These include crack propagation~\cite{Elder:2004}, layer instability and island formation~\cite{Tegze:2009,Wu:2009,Elder:2010a}), Kirkendall void formation~\cite{Elder:2011}, and the response to applied strain including uniaxial tension~\cite{Stefanovic:2006}.  The deformation of crystals under the common constraint of free boundaries, such as plane stress and uniaxial tension, while still maintaining a numerically convenient simulation domain with periodic boundary conditions is possible by using the present model with a vapor phase that has a bulk modulus significantly lower than that of the crystal.

See Supplemental Material for a full derivation of the first variations of the free energy functional, as well as a description of the numeric techniques.

\begin{acknowledgments}
The authors acknowledge helpful conversations with K.~Thornton and K.R.~Elder.  Additionally, E.J.S.~acknowledges helpful conversations with Z.T.~Trautt and Y.~Mishin, as well as K.S.~McReynolds.  {K.-A.W.} gratefully acknowledges the support of the National Science Council of Taiwan (NSC100-2112-M-007-001-MY2).  P.W.V.~is grateful for the financial support of NSF under contract DMR 1105409.
\end{acknowledgments}

\appendix

\section{Numerics and Convergence}
\label{sec:Numerics}
Equations~\ref{eqn:psiEvo}-\ref{eqn:etaEvo} are discretized and solved numerically with a semi-implicit spectral technique using a discrete time step $\Delta t$. Define
\begin{align}
	c_1 =& \left[1 + k^2 \Delta t M_\psi \left(1 - \hat{C}_2 \right)\right]^{-1},  \\
	c_2 =& -k^2 \, \Delta t \, M_\psi ,\\
	c_3 =& \left[ 1 + M_\eta \, \Delta t \left( 2 W  + \kappa k^2 \right) \right]^{-1} ,\\
	c_4 =& -M_\eta \Delta t ,\\
	N_\psi =& g \left[ \psi\left(b - 1 \right) - b\psi_v \right] \notag \\
	+ & \left(1 - g \right)\left[ - \frac{\nu}{2}\psi^2 + \frac{\xi}{3} \psi^3  \right] \notag \\
	+ & \frac{1}{2} \left[ g \left( C_2 * \psi\right) + C_2 * \left( g \psi \right) \right],\\
	N_\eta  =& 30 h \left[f_v - f_{pfc}  \right] + 2 W \eta^2 \left(2 \eta - 3 \right),
\end{align}
where $N_\psi$ and $N_\eta$ are evaluated with $\psi^n$ and $\eta^n$, the values of the fields at discrete time level $n$.  The convolution is
\begin{equation}
	C_2 * \psi = \text{IFT}\left\{ \hat{C}_2 \hat{\psi}\right\},
\end{equation}
where the hat indicates a discrete Fourier transform and $\text{IFT}\{\}$ indicates the inverse discrete Fourier transform.  With these definitions, the fields in Fourier space at time level $n+1$ are
\begin{align}
	\hat{\psi}^{n+1} & = c_1\left[ \hat{\psi}^n + c_2 \hat{N}_\psi \right], \label{eqn:psi_numeric} \\
	\hat{\eta}^{n+1} & = c_3\left[ \hat{\eta}^n + c_4 \hat{N}_\eta \right]. \label{eqn:eta_numeric}
\end{align}

\begin{figure}
\includegraphics{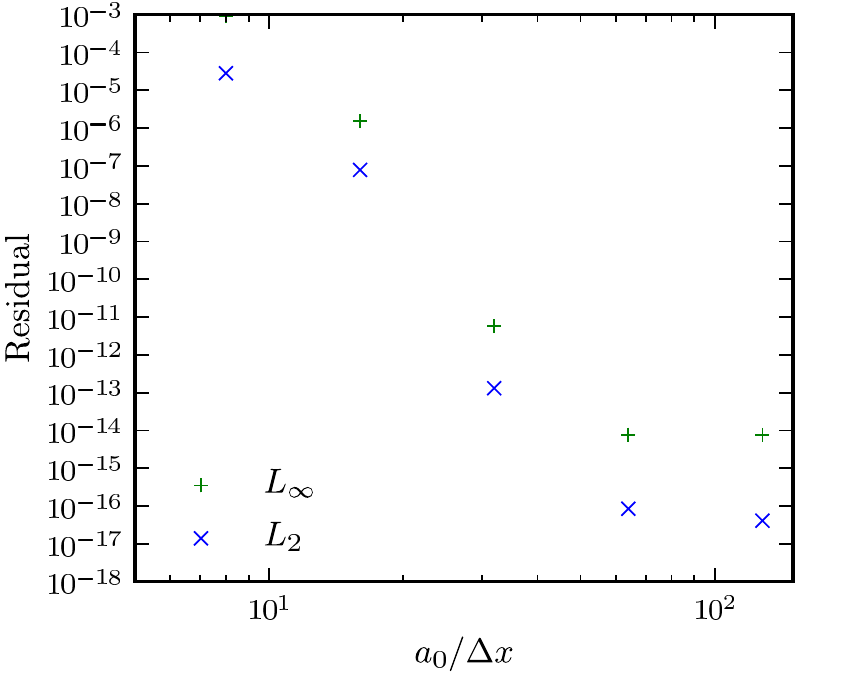}
\caption{$L_\infty$ and $L_2$ norms of the residuals as a function of the number of grid points per lattice parameter $a_0 / \Delta x$ indicating specral convergence of the density field.  Simulations parameters are from Table~\ref{tbl:HexParams}, and all numeric results in this work are reported for simulations with $a_0 / \Delta x \approx 16 $.}
\label{fig:spatialConvergence}
\end{figure}

Spatial convergence is tested by varying the grid spacing $\Delta x$ and comparing the resulting equilibrium solid density field to a reference solution.  The reference solution is computed for a single unit cell of the solid phase using a grid spacing such that $ \Delta x = a_0 / 256$, a value which is computationally intractable for large domains.  Simulations with values of $\Delta x$ as large as $a_0 / 6$ were carried out, and the residuals were computed.  The $L_\infty$ and $L_2$ norms of the residuals are shown in Fig.~\ref{fig:spatialConvergence}, which indicates spectral convergence.  Only grid points that are collocated with points on the reference solution grid are included to avoid the introduction of interpolation errors.  Simulations in this work are performed with $a_0 / \Delta x \approx 16$.  In addition to this test, the surface energy of a liquid-vapor interface was computed for systems using a similarly wide range of $\Delta x$.  Less than 1~\% change in the value of $\gamma_{lv}$ was observed over this wide range of grid spacings.

\section{Numeric Computation of the Surface Energy}
\label{sec:SurfaceEnergy}
The excess free energy of interface between phases $\alpha$ and $\beta$ within the volume $\mathcal{V}$ is given by the integral~\cite{Wu:2007}
\begin{equation}
	\gamma_{\alpha \beta} = \frac{\rho_0 k_B T}{\mathcal{A}}\int_\mathcal{V} \left[ f - \left( \bar{f}_\alpha \frac{\psi\left(\vec{r} \right) - \bar{\psi}_\beta}{\bar{\psi}_\alpha - \bar{\psi}_\beta}  - \bar{f}_\beta\frac{\psi\left(\vec{r} \right) - \bar{\psi}_\alpha}{\bar{\psi}_\alpha - \bar{\psi}_\beta} \right) \right]d\vec{r},
	\label{eqn:gammaCalculation}
\end{equation}
where $\mathcal{A}$ is the area of the interface, $f$ is the non-dimensional free energy density given in Eq.~\ref{eqn:F_functional}, and $\bar{\psi}_\nu$ and $\bar{f}_\nu$ are defined according to Eqs.~\ref{eqn:meanDensity} and \ref{eqn:MeanEnergyDensityDfn} respectively.  The domain of numeric integration for Eq.~\ref{eqn:gammaCalculation} includes both the far-field regions as well as the interface, while the integration in Eqs.~\ref{eqn:meanDensity} and \ref{eqn:MeanEnergyDensityDfn} is carried out over sub-domains of the system which are far from the interface.  

The computation of $\gamma$ is not sensitive to the choice of these regions as long as the system is in equilibrium and the sub-domains are sufficiently far from interface because the mean properties $\bar{\psi}_\nu $ and $\bar{f}_\nu$ do not vary within bulk equilibrium phases.  Finite slab thickness effects and precise choice of initial condition also introduce uncertainty into the $\gamma$ measurements.  However, we find empirically that these contributions are less than 5~\% of $\gamma$ for the choices employed in this work, consistent with the findings of Oettel et al.~\cite{Oettel:2012}.

\section{Peak fitting}
\label{sec:peakFinding}
To locate the coordinates of each peak, we first determine the coordinates of the local maximum of each of the $i$ density peaks, $\vec{r}_i = \left( x_i , y_i\right)$, by fitting a paraboloid to the data on the discrete grid in the neighborhood of each local maximum.  Let $p,q$ be the indices to the  grid points along the $x$ and $y$ directions respectively.  If the coordinates of the $i^\text{th}$ local maximum of $\rho$ on the grid occur at the location $p',q'$, the fit is performed using $\rho$ values at the grid points in the range $p=p'-1,p',p'+1, q=q'-1,q',q'+1$.  Specifically, the model for the density field is
\begin{equation}
	\rho \left( x, y \right) = a \left( x - x_i \right)^2 + b \left( y - y_i \right)^2 + c.
	\label{eqn:PeakModel}
\end{equation}
The fit parameters $x_i, y_i$ determine the center of the peak interpolated between grid points.  While this model Eq.~\ref{eqn:PeakModel} is a reasonable estimate of the peak shape within the bulk, the peaks near the solid-vapor interface, and especially near the step, are more irregularly shaped and thus there is more uncertainty in the fit.  For peaks within the bulk, the uncertainty in the coordinates reported by the fitting routine is $\sigma \approx 10^{-2} \Delta x$ where $\Delta x$ is the spacing of the discrete grid points.  Near the step, this uncertainty rises to $\sigma \approx 10^{-1} \Delta x$ due to the irregularity of the peak shapes.  The peak position uncertainty limits the $\Delta d_{n,n+1}$ measurement to an uncertainty
\begin{equation}
	\sigma_{\Delta d, n} = \sqrt{ {\sigma_n}^2 + {\sigma_{n+1}}^2}.
	\label{eqn:DispUncert}
\end{equation}
For $\sigma_n \approx \sigma_{n+1}$, we have $\sigma_{\Delta d, n} \approx \sigma_{n} \sqrt{2}$.  For our simulations $\Delta x / a_0 \approx 10^{-1}$, and thus the displacement uncertainty normalized by the lattice parameter $a_0$ ranges from $ 10^{-2}$ near the surface to $10^{-3}$ within the bulk.  Using the larger of these values, the uncertainty is not more than 10\% of the maximum displacement value near the surface, and the lower bound gives a value close to the the measured $\Delta d^\text{bulk}$.  Equation~\ref{eqn:DispUncert} is used to compute error bars in Fig.~\ref{fig:PlaneSpacing}, using the actual values of $\sigma_n$.  We note that the largest uncertainties are indeed close to the step, specifically for $n=1,w,w+1$.

\bibliography{Solid-LIquid-Vapor-PFC-Model-ArXiv}

\end{document}